\documentclass{aa} 
\input psfig.sty

\begin{document}


%
   \title{Detection of $\delta$ Scuti-like pulsation in H254, 
a pre-main sequence F-type star in IC 348\thanks{Based on 
observations collected at the Loiano Observatory, Italy and the 
San Pedro Martir Observatory, M\'exico }}


   \author{V. Ripepi\inst{1}, F. Palla\inst{2}, M. Marconi\inst{1}, 
S. Bernabei\inst{3,4}, A. Arellano Ferro\inst{5},L. Terranegra\inst{1},  
J.M. Alcal\'a\inst{1}}
  
   \offprints{}

\institute{Osservatorio Astronomico di
Capodimonte, Via Moiariello 16, I-80131 Napoli, Italy 
\and Osservatorio Astrofisico di Arcetri, Largo E. Fermi, 5, 
I-50125 Firenze, Italy 
 \and Osservatorio Astronomico di Bologna, Via Ranzani 1, 
I-40127 Bologna, Italy 
\and Departamento de Astrof{\'\i}sica, Universidad de La Laguna, 
Avda. Astrof{\'\i}sico F. S\'anchez sn, 30071 La Laguna, Spain 
\and Instituto de Astronom{\'\i}a, UNAM, Apdo. Postal 70-264, M\'exico D.F., 
CP 04510, M\'exico \\
}
\date{Received/Accepted}

\abstract{
We present time series observations of intermediate mass 
PMS stars belonging to the young star cluster IC 348. 
The new data reveal that a 
young member of the cluster, H254, undergoes periodic light 
variations with $\delta$ Scuti-like characteristics.
This occurrence provides an unambiguous evidence confirming 
the prediction that intermediate-mass pre-main sequence (PMS) 
stars should experience this transient
instability during their approach to the main-sequence. \par
On the basis of the measured frequency $f=$7.406~day$^{-1}$, we are able 
to constrain the intrinsic stellar parameters of H254 by means of 
linear, non adiabatic, radial pulsation models. The range of the 
resulting luminosity and effective temperature permitted by the
models is narrower than the observational values. 
In particular, the pulsation analysis allows to derive an independent 
estimate of the distance to IC 348 of about 320 pc.
Further observations could either confirm the monoperiodic 
nature of H254 or reveal the presence of other frequencies. 

\keywords{stars: variables:  $\delta$ Scuti  -- stars:  oscillations --
	       stars: fundamental parameters -- stars:  PMS }}

\titlerunning{Detection of $\delta$ Scuti-like pulsation in IC 348}
\authorrunning{Ripepi et al.}

   \maketitle

%

\section{Introduction}

IC~348 is a young ($\le$ 10 Myr), nearby (d$\approx$300 pc) cluster within
the Perseus complex. The cluster belongs to the Per~OB2 association, and is
located near the tip of the Perseus ridge which contains other star-forming
regions, such as NGC~1333 (Blaauw 1952; Herbig 1998).  A number of T Tauri
stars were discovered in IC~348 by Herbig (1954) who suggested that these
stars could be as young as the OB association.  The age of IC~348 has been 
debated in the literature due to the discrepancy 
between the kinematic age (1--1.4 Myr, Herbig 1998 and references therein) 
and that obtained from evolutionary considerations 
(3-20 Myr, Strom, Strom \& Carrasco 1974;
Trullols \& Jordi 1997). According to Herbig (1998), the mean age of IC 348
inferred from its faint members is much smaller than previous evolutionary
estimates, but the age spread of individual stars encompass most of
evaluations in the literature.  In fact, reconstruction of the history of 
the cluster indicates that star formation increased dramatically about $3\times
10^6$~years ago, with an $e$-folding time of the accelerating phase of
just $1\times 10^6$~yr (Palla \& Stahler 2001). Current 
estimates of the distance to IC 348 range from 240-260 pc
(Trullos \& Jordi 1997; \u{C}ernis 1993) to 316 pc (Herbig 1998). 
However, there are indications in favor of the larger distance 
(see discussion in Herbig 1998). 

Recently, about 50 new variable stars have been discovered in IC~348 by
Herbst et al. (2000) who studied their long term behavior over a period
of several months. All these variable stars are classical or weak-lined
T Tauri stars, while none of the early-type members showed any variability.
Unfortunately, this extensive work does not provide information
on the possible existence of $\delta$ Scuti-like oscillations with 
time scales of the order of hours or less in the cluster members.

During the last few years a growing interest has developed in the study of
the pulsational properties of young stars of intermediate mass (e.g., Marconi et
al. 2001, 2002; Kurtz \& Catala 2001 and references therein).  After the
initial identification of pre--main-sequence (PMS) $\delta$ Scuti candidates
in the open cluster NGC~2264 (Breger 1972), several studies
have reported results on the search for this kind of variability in known
Herbig Ae stars (Kurtz \& Marang 1995; Donati et al. 1997; Kurtz \& Muller 1999,
2001; Marconi et al.  2000, 2001; Pigulski et al. 2000; Pinheiro et al.
2002).

These investigations were stimulated by the theoretical study of
Marconi \& Palla (1998) that established the location of the instability
strip for PMS objects on the basis of nonlinear convective hydrodynamical
models.  As shown in our previous investigations (Marconi et al. 2000, 2001,
2002), the comparison between observed periodicities and the predictions of
linear non-adiabatic models provides useful constraints on the occurrence of
radial pulsations, as well as on the intrinsic stellar parameters.  Moreover,
the morphology of the PMS and post-MS tracks together with the comparison
of the predicted instability strip helps to constrain the evolutionary state
and the modal stability. 

However, apart from Breger's candidates in NGC~2264 for which only few hours
of observations are available, the other identified PMS $\delta$ Scuti are
isolated Herbig Ae stars. 
Obviously, the theoretical analysis outlined above is expected to
give the best results for stars in clusters where the uncertainties in 
distance, age and reddening are less and the same for all the members.  
For this reason, we have selected IC~348 whose small distance,
young age, and relatively large population offer a good opportunity to
investigate the short time scale pulsation of some of its members.  

The extensive study of Luhman et al. (1998) provides a detailed census of
the stellar members of the cluster. From the comparison between the stellar
parameters and the topology of the instability strip calculated by 
Marconi \& Palla (1998), we have selected three
PMS $\delta$ Scuti candidates: H83, H254, and H261 (following Herbig's 1998
notation). Their properties are listed in Table~\ref{luhman}.  
In this paper, we
present the results of an extensive observational search that resulted in the
detection of variability in one star, H254.  The paper is organized as
follows: in Section 2 observations and data analysis are discussed, whereas
the frequency analysis for the identified PMS $\delta$ Scuti candidate is
presented in Section 3. In Sect. 4 we compare the resulting periodicity with
the predictions of PMS evolutionary and pulsation models. Some final remarks
close the paper.

\begin{table*}
\caption[]{Properties of the stars H83, H254 and H261 belonging 
to IC348. Values for $T_{\rm eff}$ and $L_{\rm bol}$ are taken 
from Luhman et al. (1998). 
For the errors in $T_{\rm eff}$ and $L_{\rm bol}$ see discussion 
in Sect. 4. \label{luhman}}
\begin{tabular}{lcccccc}
\hline
\noalign{\smallskip}
Star  & $\alpha$ & $\delta$ & ST & V     & $T_{\rm eff}$ & $L_{\rm bol}$  \\
      &  (J2000) &  (J2000) &    & (mag) & (K)           & $(L_{\odot})$  \\
\noalign{\smallskip}
\hline
\noalign{\smallskip}
H83  & 3 44 19.12 & +32 09 30.8 & F0 & 11.9 & $7200\pm 170$ & $6.6\pm 2.3$  \\
H254 & 3 44 31.21 & +32 06 22.1 & F0 & 10.6 & $7200\pm 170$ & $31.4\pm 10.1$ \\
H261 & 3 44 24.67 & +32 10 14.4 & F2 & 11.6 & $6890\pm 160$ & $15.0\pm 4.8$ \\
\noalign{\smallskip}
\hline
\end{tabular}
\end{table*}

\section{Observations and data reduction}

Observations were carried out with two different telescopes: the 1.54~m 
Loiano Observatory Telescope (Bologna, Italy), and the 1.5~m telescope in 
San Pedro Martir (M\'exico, SPM hereafter). 
Observations in Loiano were obtained using both a focal reducer 
instrument (BFOSC, Bologna Faint Object Spectroscopic Camera, 
see www.bo.astro.it/loiano/observe.htm) and a three channel 
photometer (TTCP, see www.na.astro.it/$\sim$silvotti/TTCP.html and 
Ali\u{s}auskas et al. 2000). 
Observations in SPM were carried out using the six channel spectrophotometer 
(see 132.248.3.38/Instruments/danes/photomdan.html)
which can simultaneously operate the $uvby$ filters.
The observation log is reported in Table~\ref{jou}. 
 
Observations with BFOSC were obtained in the period 8-12 Sept 2001 during an
observing run dedicated to test a new set of Str\"omgren filters acquired by
the Capodimonte Astronomical Observatory.  The night of Sept 9 was lost due
to bad weather.  The BFOSC instrument is equipped with an EEV CCD
1300$\times$1340 pixel$^2$, the pixel scale is 0.58 arcsec pix$^{-1}$ for a
total field of view of 12.6$^{'} \times 13.0^{'}$.  The observations were
carried out with the Str\"omgren $by$ filters and the typical exposure times
were 40-80 s, enough to reach a very high S/N  ($\sim$ 1000) for our
candidates. In this way, we were able to obtain around 60 phase points per
night.  The seeing was bad ($\sim$ 3-4 arcsec) during the night of Sept. 8,
whereas it was around 2 arcsec or less during the other nights.  

The pre-reduction was performed in the usual 
way by subtracting a bias frame from
all the scientific frames and dividing them by a flat field (dome flat field) 
obtained during each night separately.  
Since the field of IC~348 is not crowded and the
target stars are very bright, we decided to perform aperture photometry. To
this aim, we used the DAOPHOT II program (Stetson, 1987) with a fixed circular
aperture of 40 pixels ($\sim$ 23 arcsec), whereas the sky was evaluated in an
annulus of 15 pixels around the aperture 
(starting 2 pixels beyond the central circular aperture). 
On each frame we have found and measured about 210 stars. 
In order to have compatible (instrumental) photometry in each frame 
(for a given filter), we chose a reference image taken with the best seeing 
and corrected the instrumental magnitudes of each frame by using all 
the stars in common with the reference frame. 
We note that we did not observe standard stars since we are only 
interested in the time series analysis of stars in IC348 
and for this task absolute photometry is not necessary. \par
Finally, the procedure described above allowed us to investigate
the short term variability of each candidate in IC348.

\subsection{Periodic light variability of H254}

\begin{figure}[h]
\caption{Instrumental CCD light curve of H20 and H254. \label{h20}}
\psfig{figure=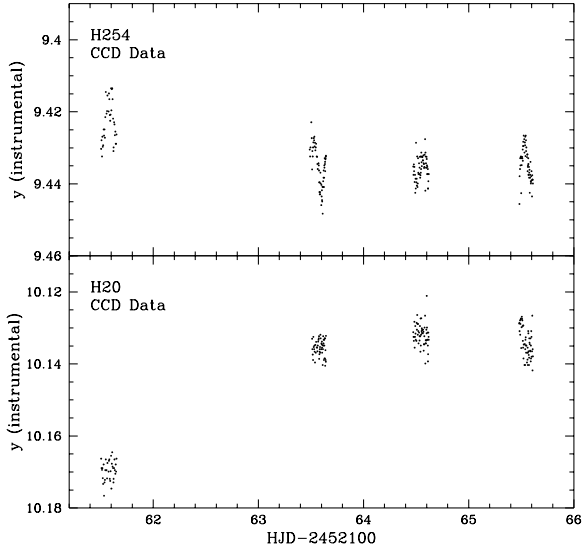,width=8cm}
\end{figure}

A plot of HJD vs. magnitude reveals that H261
and H83 are constant at a level of a few thousandths of magnitude 
on a timescale of a day, whereas they show a slight night-by-night 
variation of the mean light level of the order of 0.01 mag.
On the contrary, H254 is found to vary both in $b$ and $y$ on a timescale 
of $\sim$3 hr with an amplitude of about 0.02 mag. The light curves 
obtained in 4 nights of observations are shown in the upper panel of 
Fig.~\ref{h20}.  In addition to
fast variability, there are some indications of the presence of a slow variation
(see the difference in minimum light between first and second night 
and the small variations of the nightly means) not
correlated with the $\delta$ Scuti-like pulsation we are seeking.

This encouraging result led us to plan new observations of
H254. In particular, we aimed at using the fast three channel photometer
available at the Loiano telescope (the TTCP photometer), which has the
advantage of allowing differential photometry by observing simultaneously the
variable and comparison star and the sky.  Such a procedure guarantees
very precise photometry even in the case of thin cirrus over the sky.  Moreover,
data reduction is much easier and faster than with CCD observations.

In order to  find a suitable comparison star near H254 for 
differential photometry with the three channel photometer, we used
the available CCD observations.  The comparison star had to satisfy
the following requirements:  1) to be constant at least on the time scale of
one day; 2) to be close enough to H254 ($7^\prime<r<10^\prime$, due to the
characteristics of the TTCP); 3) to have a brightness and color
not too different from those of H254.  These constraints are reasonably
satisfied only by the star H20 at a distance of $\sim7.4^\prime$ from
H254 and of spectral type F8 (F0 for H254).  In order to verify the stability
of H20\footnote{H20 is not included in the field of Herbst et al. (2000) 
who investigated the variability in IC348 on a time scale of the order of one
day.} on timescales of less than about one day, 
we show its magnitude vs HJD in the
bottom panel of Figure~\ref{h20}. H20 is constant over periods
of several hours, whereas it appears to have night-to-night  
variations of the mean light level of about 0.03 mag. This behavior is not
surprising, considering that PMS stars can show variability due to
the interaction with circumstellar material. However, since we are interested
in short term variability only, we can use H20 as a good comparison star.

Observations of H254 with the TTCP were obtained during December 2001 (1
night) and January 2002 (6 nights) only in the $BV$ Johnson filters with
exposure times of 10 sec in both filters.

Observations in SPM were made simultaneously with the Jan. 2002
run at Loiano.  This fact guaranteed an almost continuous coverage of
H254 for about 11-12 hours on average.  The instrument used in SPM was a  
6 channel spectrophotometer which allows to observe in the Str\"omgren $uvby$ 
filters simultaneously. Since the instrument behaves as a single
channel photometer, variable and comparison were observed in the sequence
V,S,C,S...  (V=variable, S=sky, C=comparison).  Data obtained during 
Jan. 8 and 13, 2002 were rejected due to the poor sky conditions.  
On the contrary, photometric conditions occurred during the
remaining nights, allowing a calibration of H20 in terms of the standard
Str\"omgren indexes $V,b-y,m1,c1$.  We derived the following
values:  $V=11.348\pm 0.020$, $b-y=0.477\pm0.007$,
$m1=0.197\pm0.034$, $c1=0.489\pm0.061$ (average of 385 data points). 
We note that the reddening free quantities $[m1]$ and $[c1]$ 
(Str\"omgren, 1966) for H20 are consistent with its spectral type.

Before combining the three photometric datasets, one has to take 
into account that they
were obtained in two different photometric systems. 
Johnson $V$ and Str\"omgren $y$ are easier and safer to combine than
Johnson $B$ and Str\"omgren $b$, because they differ only by a
constant and a slight color term in $b-y$ (see e.g. Terranegra et al. 1994). 
The former is removed when one
uses differential photometry and the latter can be neglected since H20 and
H254 are both F-type stars.
On this basis, $\delta y \approx \delta V$  and in the following we shall
always refer to the data in the $V$ band.  The last step before proceeding to
the frequency analysis was to rebin the Loiano photometry obtained with TTCP
(averaging HJD and photometry every 5 data points) in order to have similar
sampling and, in turn, a similar weight for each dataset in the fitting
procedure. \par
\begin{figure*}
\caption{Differential light curve for H254 
($\delta V$ = $V_{H254}-V_{H20}$). CCD data are for HJD from 61 to 66,  
while the remaining data are those obtained with TTCP at Loiano and 
6 chan. spect. at SPM.\label{phot}}
\psfig{figure=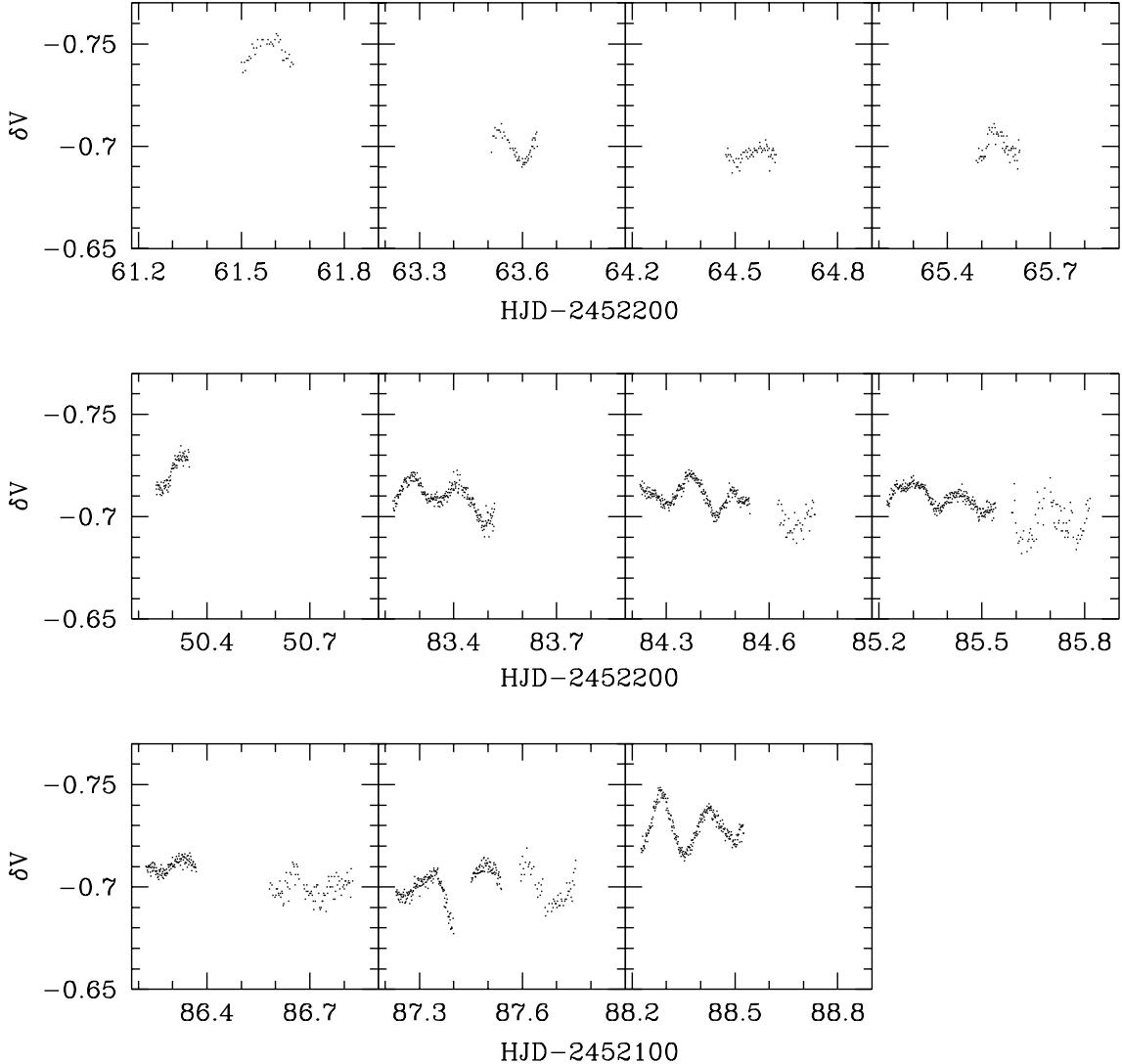,width=16cm}
\end{figure*}

 
The final dataset is shown in Figure~\ref{phot}. We note that
in this figure the data between nights 2452161-2452165 consist of differential
photometry ($\delta V$ = $V_{H254}-V_{H20}$) obtained from CCD observations. A
comparison between these data and the ones shown in the bottom panel of
Fig.~\ref{h20} reveals that using H20 as comparison star leads to 
 more precise photometry.

\begin{table}
\caption[]{Journal of the observations: $uvby$ are in the Str\"omgren 
system, $BV$ are in the Johnson system. \label{jou}}
\begin{tabular}{lllll}
\hline
\noalign{\smallskip}
Period  & Teles. & Instr. & Filt. & Star  \\
\noalign{\smallskip}
\hline
\noalign{\smallskip}
8,10$-$12/9/01 & Loiano & BFOSC    & $by$   & H83,H254,H261   \\
6/12/01        & Loiano & TTCP     & $BV$   & H254            \\
8$-$13/1/02    & Loiano & TTCP     & $BV$   & H254            \\
8$-$13/1/02    & SPM    & 6 ch. sp.   & $uvby$ & H254            \\
\noalign{\smallskip}
\hline
\end{tabular}
\end{table}

\section{Frequency analysis of H254}

\begin{figure}
\caption{Spectral window of our data set. \label{fig2}}
\psfig{figure=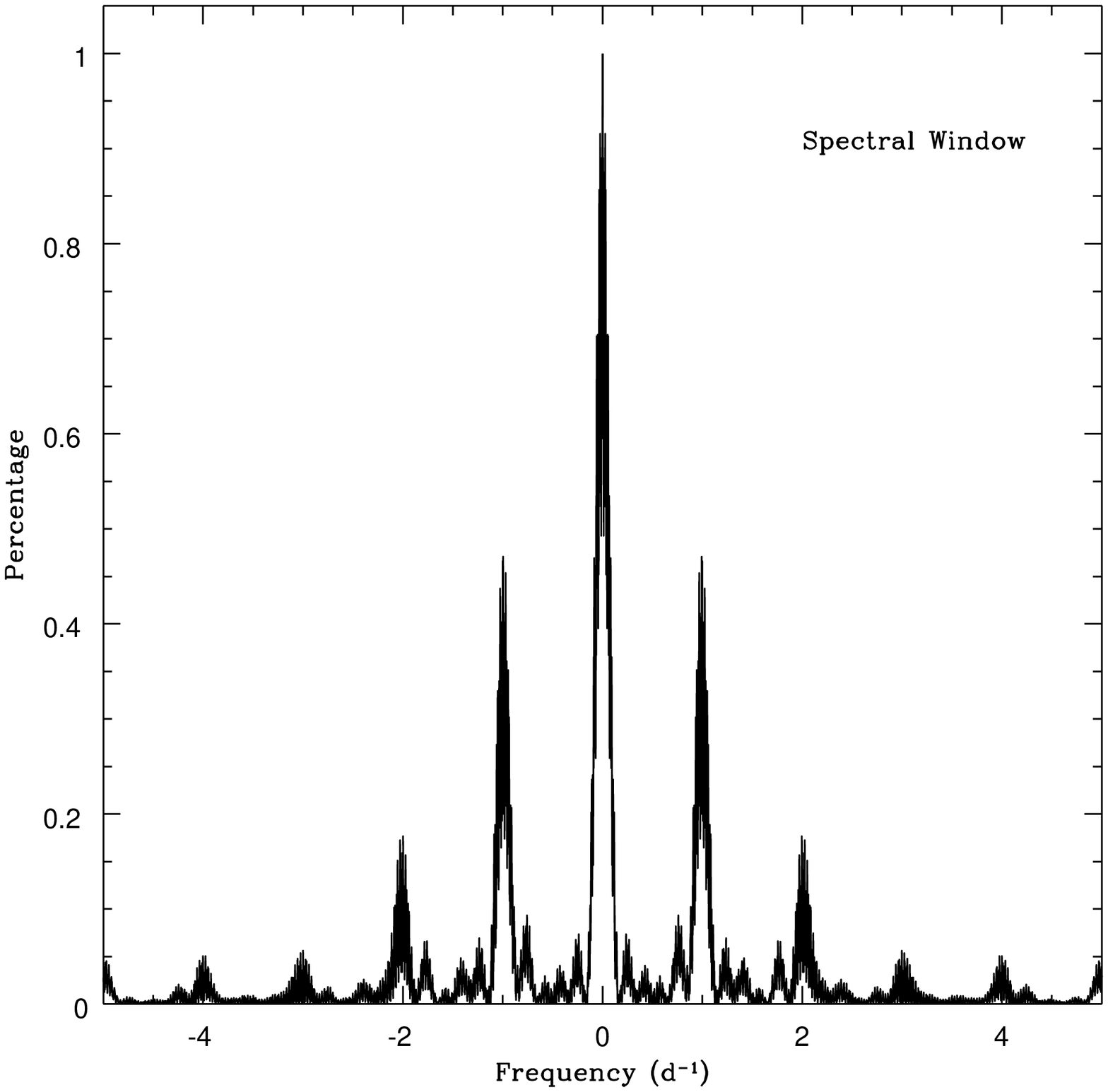,height=7cm}
\end{figure}

The frequency analysis was performed using the Period98 package
(available at www.astro.univie.ac.at/$\sim$dsn/), based on the
Fourier transform method. For a better
interpretation of the results, we have first calculated the spectral window 
for the whole data set. The result is shown in Fig.~\ref{fig2}: any good
frequency should display the same alias pattern as that shown in the figure.

\begin{figure}
\caption{Periodograms of $V$ photometry for H254. The sequence from top to
bottom shows the change of the periodogram pre-whitened by $f_1$, $f_2$,
$f_3$, and $f_4$, respectively.\label{fig3}}
\psfig{figure=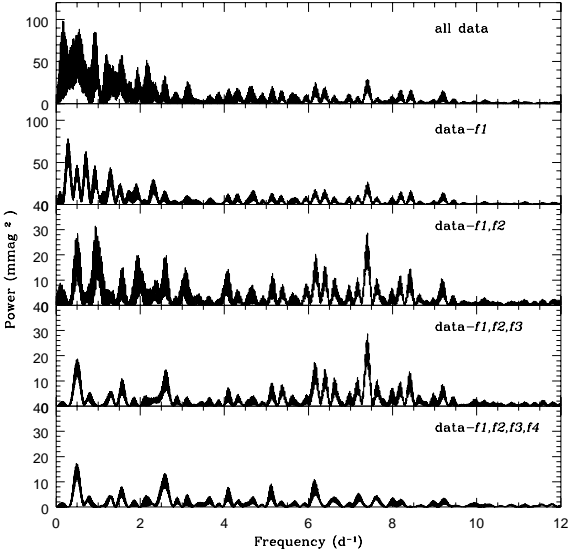,width=8cm}
\end{figure}

\begin{figure}
\caption{Light curve of H254 obtained using the frequency 
$f_4$=7.406~day$^{-1}$ and after removal of the frequencies $f_1,f_2,f_3$. 
The solid line shows the same data after averaging magnitudes 
over a 10 phase bin. \label{fasa}}
\psfig{figure=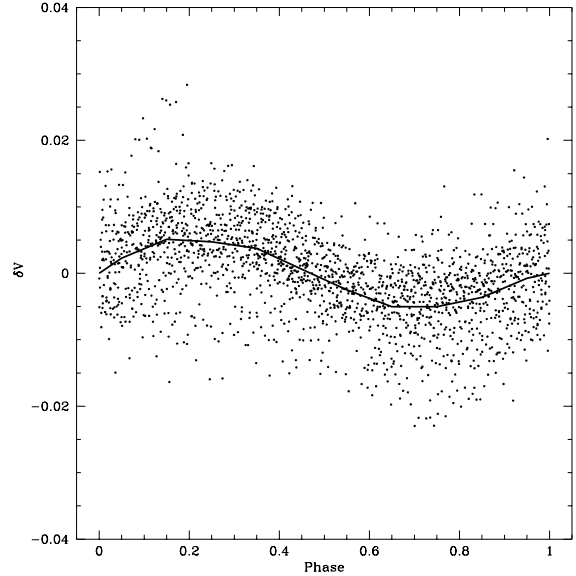,width=8cm}
\end{figure}

As seen in Figure~\ref{fig3}, the Fourier analysis identifies
four frequencies, whose properties are listed in Table~\ref{tab2}.  The low
frequencies with $f_1$=0.157, $f_2$=0.283, and $f_3$=0.931~day$^{-1}$ result
from the long term behavior associated with a daily variation of H254 and,
partially, with the similar variability in the comparison star, H20. After
removal of $f_1$, $f_2$~and $f_3$, a fourth frequency is identified at
$f_4$=7.406~day$^{-1}$ (3.24~hr) which is typical of 
$\delta$ Scuti type pulsators.  After the pre-whitening of $f_4$, no other
significant features are seen in the periodograms.
In order to check the accuracy of our analysis, we show in Figure~\ref{fasa}
the light curve phased with the frequency $f_4$=7.406~day$^{-1}$ after 
pre-whitening for the other frequencies. The figure reveals that almost all 
the data can be satisfactorily phased when using the quoted frequency.\par
We must caution that the presence of a one day alias makes our determination
of the frequency still not definitive. In particular, the $-$1~day alias
(i.e., a frequency $\sim$6.4~day$^{-1}$) shows an excess of power with
respect to that expected on the basis of the spectral window, suggesting that
the true frequency could be the latter. This must be taken into account when
interpreting the data in terms of the predictions of the pulsation models.

\begin{table}
\caption[]{Frequencies, amplitudes and phases derived for the Fourier 
analysis of the data. The uncertainty on the frequencies is 
$\sim$0.008~d$^{-1}$
\label{tab2}}
\begin{tabular}{llll}
\hline
\noalign{\smallskip}
  & Frequency   &   Amplitude & Phase  \\
  & (d$^{-1}$)  & (mmag)      &   \\
\noalign{\smallskip}
\hline
\noalign{\smallskip}
{\it $f_1$} & 0.157  & 12.5 & 0.001   \\
{\it $f_2$} & 0.283  &  9.0 & 0.758   \\
{\it $f_3$} & 0.931  &  6.5 & 0.065   \\
{\it $f_4$} & 7.406  &  5.4 & 0.028   \\
\noalign{\smallskip}
\hline
\end{tabular}
\end{table}

\section{Theoretical constraints}

\begin{figure}
\caption{Location of H83, H254 and H261 in the HR diagram. The filled circles
represent the three $\delta$ Scuti candidates, while the 
empty circles represent 
the intermediate-mass cluster members. Stellar parameters are taken
from Luhman et al. (1998). Also shown are the evolutionary
tracks and isochrones by Palla \& Stahler (1993), and the
instability strip by Marconi \& Palla (1998).   \label{strip}}
\psfig{figure=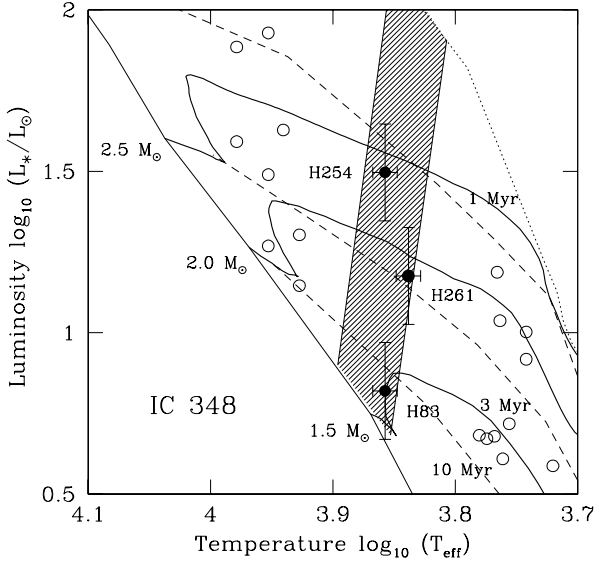,width=8cm}
\end{figure}

Figure~\ref{strip} shows the location in the HR 
diagram of the three PMS $\delta$ Scuti candidates in IC~348,
along with several other cluster members of intermediate mass.
The physical parameters are taken from Luhman et al. (1998) who assumed 
a distance to IC 348 of 316 pc (i.e. DM=7.5 mag) as suggested by 
Herbig (1998).  
Unfortunately Luhman et al. do not provide the error estimate 
on effective temperature and luminosity. 
The values given in Table~\ref{luhman}, and reported in 
Fig.~\ref{strip}, represent our estimates obtained by 
propagating the errors due 
to the adopted calibrations between spectral type and T$_{\rm eff}$ 
 and T$_{\rm eff}$ vs. bolometric correction. In particular, 
the error bar in the effective temperature comes from the uncertainty
of $\pm$ 1 spectral class (using the calibrations of Schmidt-Kaler 1982,
and Leggett et al. 1996), while the error on the luminosity 
comes mainly from the uncertainty in the bolometric correction
when propagating the error in T$_{\rm eff}$ (using the relation 
between T$_{\rm eff}$, intrinsic colors and bolometric corrections of 
Kenyon and Hartmann 1995). \par

The cluster contains a non negligible population of stars with mass greater 
than $\sim$1.5 M$_\odot$. However, only three objects fall within the
boundaries of the theoretical instability strip calculated 
by Marconi \& Palla (1998). 
Interestingly, their estimated ages cover a large range 
from $\sim$1~Myr for H254 to $\sim$10~Myr
for H83, as can be deduced from the isochrones shown in the HR diagram.
The three candidates lie within the instability 
region, with H261 and H83 located close to the red 
edge where pulsation amplitudes are expected to be very small.
The theoretical uncertainty on the red boundary is
of the order of $\pm$0.02~dex in $\log T_{\rm eff}$ due to the sensitivity on 
the assumed mixing length parameter in the treatment of convection 
(see Stellingwerf 1982 for details on the treatment of convection in the 
pulsation code) which is more efficient
toward the cooler part of the strip\footnote{Due to the lower efficiency of 
convection at higher effective temperatures, the uncertainty on the 
location of the blue boundary of the strip is about $\pm$0.005 dex.}.
This is consistent with the fact that a $\delta$~Scuti type 
variation has been found only in H254.
Conversely, the lack of pulsation provides a strong empirical constraint 
to the red limit of the instability strip.   

As shown in Fig.~\ref{strip}, H254 is located well inside the strip
in a region where pulsation in the lowest modes is expected.
In order to reproduce the observed periodicity of 7.406~day$^{-1}$, 
we have computed a sequence of linear non-adiabatic models with
solar metallicity, varying the effective temperature and luminosity
within the ranges given above. The stellar mass is evaluated from
the evolutionary tracks. As a result, we find that 
H254 pulsates either in the fundamental mode or in the first overtone. 
Higher modes seem to be ruled out by the lack of consistency
with the observed luminosity and effective temperature ranges and 
the observed periodicity.  
The values of mass, effective temperature and luminosity that 
characterize these models are listed in Table~\ref{tab3}.
In the table, we have also considered the frequency 6.406 d$^{-1}$ that
yields values of mass and luminosity in between those for the 7.406~day$^{-1}$. 
Note that in any case the range of stellar parameters permitted by the
models is narrower than the observational values. 

Unfortunately, the uncertainties on luminosity and effective temperature 
still preclude a secure identification of the pulsation mode.  Future
higher quality observations of the short term photometric behavior
will help to resolve the exact pulsation properties of H254. However,
since we could not reproduce the observed period with any other
combination of input parameters, our results clearly indicate that the
pulsation period of 7.406~day$^{-1}$ (or 6.406 d$^{-1}$) is
incompatible with the possibility that H254 is of lower luminosity
(hence mass) as in the case of a shorter distance, namely 240-260 pc 
(Trullos \& Jordi 1997; \u{C}ernis 1993) instead of 316 pc (Herbig 1998).

\begin{figure}
\caption{Enlargement of the HR diagram for H254. 
The other symbols represent the best fit radial pulsation models with the
values of the stellar parameters given in Table~\ref{tab3}: squares are
for the 7.406~day$^{-1}$~period, while the triangle corresponds to 
the $-1$d alias.  \label{strip_h254}}
\psfig{figure=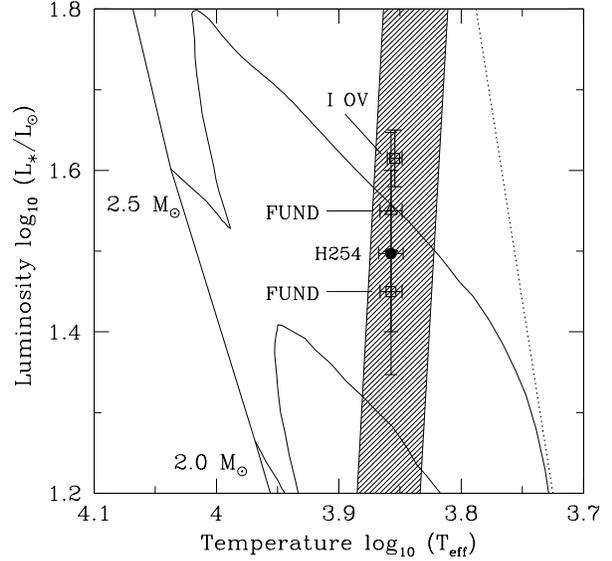,width=8cm}
\end{figure}

\begin{table}
\caption[]{Best fit of radial pulsation models.  \label{tab3}}
\begin{tabular}{lcccc}
\hline
\noalign{\smallskip}
Period      &   Mass      & $T_{\rm eff}$ & log L       & Mode        \\
($d^{-1}$)  & ($M_\odot$) & (K)           & ($L_\odot$) &             \\
\noalign{\smallskip}
\hline
\noalign{\smallskip}
7.406 & $2.6\pm 0.1$   & $7150\pm 100$ & $1.62\pm 0.04$ & I OV \\
7.406 & $2.3\pm 0.1$   & $7200\pm 150$ & $1.45\pm 0.05$ & FUND \\
6.406 & $2.50\pm 0.05$ & $7200\pm 150$ & $1.55\pm 0.05$ & FUND \\
\noalign{\smallskip}
\hline
\end{tabular}
\end{table}

\section{Conclusions}

The present observations have revealed that the PMS F-type star H254, 
member of IC~348, undergoes periodic light variations with 
$\delta$ Scuti-like frequency.
Since IC 348 has an estimated age of less than $\sim$10 Myr and is still
actively forming stars, the discovery of small amplitude pulsations in H254
confirms the prediction by Marconi \& Palla (1998) 
that intermediate-mass PMS stars should experience this transient
instability during their approach to the main-sequence. Other cases presented 
in the literature have some ambiguity regarding the actual evolutionary 
state, i.e. pre--main- vs. post--main-sequence (e.g., V351 Ori 
described in Marconi et al. 2001). Thus, IC~348 and NGC~2264 are 
the only two young clusters where $\delta$ Scuti-like  pulsations 
in PMS stars have been detected so far. 
Similar searches should be conducted on other clusters to
enlarge the sample of pulsating intermediate-mass stars. Good candidates can
be found in, e.g., NGC~2362 (Moitinho et al. 2001) and the Upper Sco-Cen
association (Preibisch \& Zinnecker 1999) that contain a rather large
population of stars of the appropriate spectral types.

In H254, we find only one mode of pulsation with
$f_4=$7.406~day$^{-1}$ (or 3.24 hr). Non-adiabatic linear models with the
observed parameters of H254 show unstable modes of low order, namely the
fundamental or the first overtone mode. The occurrence of just one mode
of pulsations could be intrinsic or due to the detection threshold.
Theoretical models of classical $\delta $ Scuti stars predict many
unstable modes, in excess of what is actually observed (e.g. Bradley \&
Guzik 2000). On the other hand, monoperiodic radially pulsating $\delta$
Scuti stars are known (20~CVn being the best example, Chadid et al. 2001), 
and H254 may represent the young counterpart of this (admittedly limited)
class. Spectroscopic observations of this star should be carried out to
confirm the photometric period found by us and to verify whether other
modes are also present. Of course, the low amplitude of the modes may
render the detection quite difficult. However, the identification of a
young, monoperiodic pulsating star will be very useful to shed light 
on the physical mechanism that limits the amplitude of the pulsations.
Thus, we encourage other groups interested in stellar pulsations to consider 
H254 for further study. 

\begin{acknowledgements}
We are very grateful to our referee, Dr. W. Herbst, for his valuable 
comments and suggestions.
We wish to thank the Loiano and San Pedro Martir Observatories 
staff members for their kind support during the observations. 
Partial financial support for this work was provided 
by MURST-Cofin 1998, under the project 
``Physics of solar and stellar external atmospheres along 
their evolutionary tracks''.  
This work made use of the SIMBAD database at CDS.
\end{acknowledgements}

\end{document}